\documentclass[prb,aps,twocolumn,showpacs]{revtex4}
\usepackage{epsfig}
\include{psfig}

\begin{document}


\title{\bf Energy dependent counting statistics in diffusive superconducting tunnel junctions.}
\author{P. Samuelsson}
\affiliation{D\'epartement de Physique Th\'eorique, Universit\'e de
Gen\`eve, CH-1211 Gen\`eve 4, Switzerland.}  
\date{\today}
\begin{abstract}
We present an investigation of the energy dependence of the full
charge counting statistics in diffusive
normal-insulating-normal-insulating-superconducting junctions. It is
found that the current in general is transported via a correlated
transfer of pairs of electrons. Only in the case of strongly
asymmetric tunnel barriers or energies much larger than the Thouless
energy is the pair transfer uncorrelated. The second cumulant, the
noise, is found to depend strongly on the applied voltage and
temperature. For a junction resistance dominated by the tunnel barrier
to the normal reservoir, the differential shot noise shows a double
peak feature at voltages of the order of the Thouless energy, a
signature of an ensemble averaged electron-hole resonance.
\end{abstract}

\maketitle

The mechanism for current transport across a normal
conductor-superconducting (NS) interface, at energies below the
superconducting gap, is Andreev reflection. An electron incident from
the normal conductor is retroreflected into a hole, and a charge of
$2e$ is transported into the superconductor. The fact that electrons
are transfered across the NS-interface in pairs has recently attracted
a lot of interest to noise in NS-systems, \cite{Buttikerrew} since the
noise can provide information about the charge transfer mechanism.

In some NS-systems, the shot noise is doubled compared to the noise in
the corresponding normal system. In a
normal-insulating-superconducting (NIS) junction, the pairs tunnel
with a small probability between the normal and the superconducting
reservoirs. This gives rise to a Poisson noise of pairs of electrons,
which is thus twice as large as standard Poisson
noise. \cite{Khlus87,deJong94,Muz94} Such pair Poissonian noise was
recently observed experimentally. \cite{Lefloch}

Also in diffusive wire NS-junctions, a doubling of the shot noise has
been predicted theoretically \cite{deJong94,Nagaev01,Belzig01a} and
observed experimentally \cite{Jehl00,Koz00,Jehl01} for applied
voltages much smaller as well as much larger than the Thouless
energy. However, for intermediate voltages, of the order of the
Thouless energy, the shot noise deviates from the double normal value,
\cite{Belzig01a,Reulet02} a behavior related to the induced proximity
effect in the normal conductor.

In general, in mesoscopic NS-systems, there is no simple relation
between the shot noise in the normal and the superconducting
state. This has been demonstrated for a wide variety of multi-mode
systems such as diffusive NIS-junctions, \cite{deJong94,Hekkila02}
chaotic-dot superconductor junctions \cite{deJong97,Samuelsson02a},
multiterminal superconducting tunnel junctions\cite{Boerlin02} and
wide ballistic double-barrier\cite{deJong97,Fauchere98} and normal
slab\cite{Schechter01} superconductor junctions, as well as few-mode
superconducting systems containing
beam-splitters,\cite{Datta96,Martin96,Torres99,Taddei02} disordered
tunnel junctions \cite{Gramespacher99} and double barrier
junctions.\cite{Khlus95}
 
To obtain complete knowledge of the charge transfer in NS-systems, it
is thus not sufficient to study the shot noise. Instead, one has to
investigate the full charge counting statistics.\cite{Levitov} This
was originally done by Muzykanskii and Khmelnitskii,\cite{Muz94} who
showed that the statistics of the charge transfer across a
NS-interface with arbitrary transparency, can in general be described
as a correlated transfer of pairs of electrons. Only in the limit of a
low transparency $NIS$ junction, the transfer of pairs is
uncorrelated, Poissonian. Using a recently developed circuit theory
approach \cite{Belzig01a,Belzig01b,Boerlin02,Belzig02} to full
counting statistics, the charge transfer statistics has been studied
in several systems containing superconductors, such as short
SNS-junctions, \cite{Belzig01b} diffusive
NS-junctions,\cite{Belzig01a} diffusive superconducting tunnel
junctions\cite{Boerlin02} and chaotic dot-superconducing
junctions. \cite{Samuelsson02b} However, in these works, the main
focus has been on the low voltage- and temperature properties, the
full energy dependence of the counting statistics was only analyzed
numerically in Ref. [\onlinecite{Belzig01a}].

In this paper we investigate the energy dependent full counting
statistics of a diffusive
normal-insulating-normal-insulating-superconducting ($NI_1N'I_2S$)
junction. This system is interesting for several reasons. Using the
circuit theory approach, \cite{Belzig01a,Belzig01b,Boerlin02} an
analytical expression for the full voltage and temperature dependence
of the counting statistics can be derived. This allows us to
investigate in detail the energy dependence of the charge transfer
mechanism. We focus on the role of the proximity effect, which is of
particular interest, since it gives rise to a large, energy dependent
modification of the transport properties. Moreover, the conductance in
$NI_1N'I_2S$ systems was recently studied
experimentally,\cite{Lefloch02} and very good agreement was found
between the experiment and the quasiclassical Greens function
theory\cite{Volkov} on which the circuit theory is based. This makes a
detailed analysis of the experimentally accessible noise of interest.

We find that the current is in general transported via a correlated
transfer of pairs of electrons. However, in the cases where the
resistance is dominated by the tunnel barrier $I_2$, at the $N'I_2S$
interface, or for energies well above the Thouless energy, the pair
transfer is uncorrelated. These cases correspond to the limit of weak
proximity effect in the normal region $N'$. In the opposite regime of
strong proximity effect in $N'$, when the resistance is dominated by
the tunnel barrier $I_1$ at the $NI_2N'$ interface, a proximity gap
opens up in the $N'$-region. For energies below the induced proximity
gap, the junction is effectively a NIS-junction and the pair transfer
is again uncorrelated.

The second cumulant, the noise, is found to depend strongly on the
applied voltage and temperature. For a junction resistance dominated
by the tunnel barrier $I_1$, at the $NI_1N'$ interface, and a
temperature well below the Thouless energy, the differential noise
shows a double-peak feature at voltages of the order of the Thouless
energy. We show that this double-peak behavior can be attributed to an
ensemble averaged electron-hole resonance.

The paper is organized as follows. We first present the circuit theory
model and the derivation of the full charge counting
statistics. Thereafter, the dependence of the counting statistics on
energy and tunnel barrier ($I_1$ and $I_2$) conductances is
analyzed. We then investigate the temperature and voltage dependence
of the noise in detail.

\section{Model and theory}

\begin{figure}[h]        
\centerline{\psfig{figure=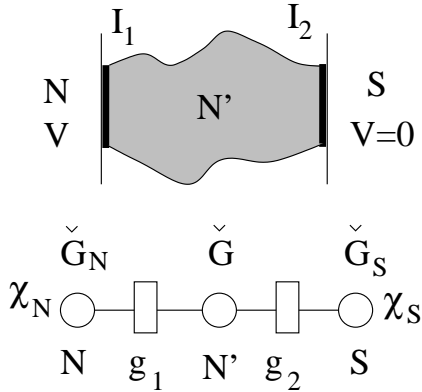,width=5.5cm}}
\caption{The $NI_1N'I_2S$-junction. Upper figure: A normal ($N$) and a
superconducting ($S$) reservoir are connected via a diffusive normal
metal ($N'$, shaded) contact. At the $NN'$ and $N'S$ interface, there
are tunnel barriers $I_1$ and $I_2$ with conductance $g_1$ and $g_2$
respectively. Lower figure: The circuit theory representation of the
junction, with the matrix Greens functions $\check G_N$, $\check G_S$
and $\check G$ and the counting fields $\chi_N$ and $\chi_S$ shown.}
\label{fig1}
\end{figure}

We consider a diffusive
normal-insulating-normal-insulating-superconducting ($NI_1N'I_2S$)
junction (see Fig. \ref{fig1}), where both the resistances of the
insulating tunnel barriers ($I_1$ and $I_2$) are much larger than the
resistance of the normal diffusive region ($N'$) between the
barriers. The short diffusive region $N'$, which does not contribute
to the junction resistance, leads to complete isotropization of the
system parameters between the barriers. \cite{Nazcirc} This makes the
junction effectively zero dimensional \cite{dotcom} and the exact
shape and size of the diffusive region is thus not important.

The conductances $g_1$ and $g_2$ of the two tunnel barriers can be
written as $g_{1(2)}=(2e^2/h)\sum_{n=1}^{N_{1(2)}}\Gamma_{1(2),n}$,
where $N_{1(2)}$ is the number of transport modes at the barrier
$1(2)$ and $\Gamma_{1(2),n}$ is the transparency of mode $n$. We
consider the limit of large number of modes, $N_{1(2)} \gg 1$ and
small transparency of each mode, $\Gamma_{1(2),n}\ll 1$. However, the
conductances $g_{1}$ and $g_2$ are much larger than the conductance
quanta $2e^2/h$, i.e. $\sum_{n=1}^{N_{1(2)}}\Gamma_{1(2),n} \gg
1$. Under these conditions we can neglect Coulomb blockade effects as
well as weak localization like corrections to the cumulants of the
current.

It is further assumed that the dwell time of the particles in the the
normal region $N'$ is much shorter than the inelastic scattering- and
phase breaking times. We are also only interested in zero frequency
properties.\cite{spin} With these assumptions, we can analyze the full
counting statistics of the junction within the framework of the
circuit theory, recently developed by Nazarov\cite{Nazcirc,Nazgenfcn}
and Belzig and Nazarov.\cite{Belzig01a,Belzig01b} Very recently, a
detailed discussion of the circuit theory was given. \cite{Belzig02}
We note that the zero energy counting statistics was studied for the
same junction in Ref. [\onlinecite{Boerlin02}], here we study the full
energy dependence.

In the circuit theory, the mesoscopic conductor is described in terms
of nodes connected via various circuit elements, just as in an
ordinary electric circuit theory. Due to the effective
zero-dimensionality of the junction, the circuit theory representation
(see Fig. \ref{fig1}) only contains three nodes, the two reservoirs
($N$ and $S$) and the isotropic normal region ($N'$) between the
tunnel barriers, and two circuit elements, the tunnel barriers ($I_1$
and $I_2$).

However, instead of characterizing each node with an electric
potential, as is done in an ordinary electrical circuit theory, they
are characterized with a $4\times 4$ matrix Green's function (in
Keldysh-Nambu space). Here we have the three node Green's functions
$\check G_N, \check G_S$ and $\check G$, as shown in Fig. \ref{fig1}.
They obey the normalization condition $\check G_N^2=\check
G_S^2=\check G^2=\check 1$. Throughout the paper we use the convention
with ``check'', $\check.$ denoting $4\times4$ matrices and ``hat''
$\hat.$ denoting $2\times2$ matrices.

The matrices of the reservoir nodes are known. In the normal reservoir
we have
\begin{eqnarray}
\check G_N&=&e^{i\chi_N\check \tau_K/2}\check G_N^0e^{-i\chi_N\check
\tau_K/2},~\check \tau_K=\left(\begin{array}{cc} 0 & \hat\sigma_z \\
\hat \sigma_z & 0 \end{array}\right) \\ \check
G_N^0&=&\left(\begin{array}{cc} \hat \sigma_z & \hat K_N \\ 0 &
-\hat\sigma_z \end{array}\right),~\hat K_N=2\left(\begin{array}{cc} 1-2f_+ & 0 \\ 0 & 2f_--1 \end{array}\right), \nonumber
\label{nres}
\end{eqnarray}
where $f_{\pm}(E)=f(E\pm eV)$, with $f(E)=(\mbox{exp}[E/kT]+1)^{-1}$
the Fermi distribution function. Here $\chi_N$ is a counting field,
``counting'' the electrons passing in and out of the normal reservoir
and $\hat \sigma_z$ is a Pauli matrix. In the superconducting reservoir we
have
\begin{eqnarray}
\check G_S&=&e^{i\chi_S\check \tau_K/2}\check G_S^0e^{-i\chi_S\check
\tau_K/2} \nonumber \\
\check G_S^0&=&\left(\begin{array}{cc} \hat R_S & \hat K_S \\ 0 & \hat A_S \end{array}\right),~\hat R(\hat A)_S=\left(\begin{array}{cc} g_{R(A)} & f_{R(A)} \\ f_{R(A)} & -g_{R(A)} \end{array}\right), \nonumber \\  
~\hat K_S&=&(\hat R_S-\hat A_S)[f(E)-f(-E)],
\label{sres}
\end{eqnarray}
where $\Delta$ is the superconducting gap and $f_{R(A)}=i\Delta[(E\pm
i\epsilon)^2-\Delta^2]^{-1/2}$ and $g_{R(A)}=[1-f_{R(A)}^2]^{1/2}$ are
Green's functions characterizing the superconducting reservoir
($\epsilon$ is an infinitesimal positive number). The counting field
$\chi_S$ of the superconducting reservoir is introduced for notational
clarity, due to current conservation it is in principle not needed.

The matrix $\check G$ of the node between the tunnel barriers is to be
determined. It is found from a conservation law for matrix currents (a
matrix Kirchoff's rule), as
\begin{eqnarray}
\check I_1+\check I_2+\check I_{E}=0
\label{curreq}
\end{eqnarray}
where the matrix currents are given by
\begin{eqnarray}
\check I_1&=&\frac{g_1}{2}[\check G_N,\check G],~ \check I_2=\frac{g_2}{2}[\check G_S,\check G] \nonumber \\
\check I_{E}&=&-\frac{2iE}{\delta}[\check G_{E},\check G],~ \check G_{E}=\left(\begin{array}{cc} \hat \sigma_z & 0 \\ 0 & \hat \sigma_z \end{array}\right)
\label{matcurr}
\end{eqnarray}
where $[A,B]=AB-BA$ and $\delta$ is the mean level spacing in the
$N'$-region divided by the conductance quanta $2e^2/h$. The Thouless
energy is given by $E_{Th}=(g_1+g_2)\delta/4$, i.e. it depends only on
the escape time through the tunnel barriers, not on the diffusion time
through the $N'$-region. The term $\check I_{E}$ leads to a
modification of the proximity effect at finite energies, and
eventually a suppression for $E \gg E_{Th}$.

From Eq. (\ref{curreq}) and the condition $\check G^2=1$, we can
determine the missing node matrix $\check G(\chi_N,\chi_S)$ as a
function of the counting fields. Following the lines of
Ref. [\onlinecite{Boerlin02}], noting that Eq. (\ref{curreq}) can be
written in the form $[\check G,\check G_x]=0$, we find the physically
relevant solution\cite{relevant}
\begin{eqnarray}
\check G&=&\check G_x(\check G_x^2)^{-1/2}, \nonumber \\
\check G_x&=&\frac{g_1}{2}\check G_N+\frac{g_2}{2}\check G_S-i\frac{E}{2E_{Th}}(g_1+g_2)\check G_{E}.
\label{gexpr}
\end{eqnarray}
Knowing $\check G(\chi_N,\chi_S)$, we can find the cumulant generating
function $F(\chi_N,\chi_S)$. It is given from the general relation
between the matrix current and the generating
function,\cite{Nazgenfcn}
\begin{eqnarray}
\frac{\partial F(\chi_N,\chi_S)}{\partial \chi_{N}}&=&\frac{it_0}{8e^2}\int dE~\mbox{tr}\left[\check \tau_K \check I_{1}(\chi_N,\chi_S)\right] \nonumber \\
\frac{\partial F(\chi_N,\chi_S)}{\partial \chi_{S}}&=&\frac{it_0}{8e^2}\int dE~\mbox{tr}\left[\check \tau_K \check I_{2}(\chi_N,\chi_S)\right] 
\end{eqnarray}
where $t_0$ is the measurement time. Using that under the trace,
\begin{eqnarray}
\frac{\partial (\check G_x^2)^{1/2}}{\partial
\chi_{N(S)}}&=&\frac{\partial \check G_x}{\partial \chi_{N(S)}}\check
G_x(\check G_x^2)^{-1/2}=\frac{i}{4}g_{1(2)}[\check \tau_K,\check
G_{N(S)}]\check G \nonumber \\ &=&\frac{i}{4}g_{1(2)}\check \tau_K
[\check G_{N(S)},\check G]=\frac{i}{2}\check \tau_KI_{1(2)},
\end{eqnarray}
we get the counting field dependent part of the
cumulant generating function
\begin{eqnarray}
F(\chi_N,\chi_S)&=&\frac{t_0}{4e^2}\int dE~\mbox{tr}\left[(\check G_x^2)^{1/2}\right].
\label{cgf}
\end{eqnarray}
We note that the standard definition of a generating function demands
$F(\chi_N=0,\chi_S=0)=0$. This implies that a counting field
independent constant $-\mbox{tr}[G_x^2(\chi_N=0,\chi_S=0)]^{1/2}$
should in principle be added to the integrand in
Eq. (\ref{cgf}). However, such a constant does not alter the
cumulants, and below we will for simplicity omit it.

The probability $P(Q)$ that $Q$ electron charges have been transported
through the junction in time $t_0$ is given by 
\begin{eqnarray}
P(Q)&=&\int_{-\pi}^{\pi}~\frac{d\chi_N}{2\pi}~e^{-i\chi_NQ-F(\chi_N,\chi_S=0)}
\label{qcount}
\end{eqnarray}
Eqs. (\ref{nres}) - (\ref{qcount}) completely determine the full
charge counting statistics for arbitrary temperatures and
voltages. For our purposes, it is however more convenient to focus on
the properties of the cumulant generating function in
Eq. (\ref{cgf}). From the cumulant generating function we can derive
all cumulants of the current by repeatedly taking the derivative with
respect to the counting field $\chi_N$ (or $\chi_S$). For the two
first cumulants, current and noise, we have (evaluated at the
$N$-contact),
\begin{eqnarray}
I&=&i\frac{e}{t_0}\left.\frac{\partial F(\chi_N,\chi_S)}{\partial \chi_N}\right|_{\chi_N=\chi_S=0}, 
\label{curr}
\end{eqnarray}
\begin{eqnarray}
S&=&-\frac{2e^2}{t_0}\left.\frac{\partial^2F(\chi_N,\chi_S)}{\partial
\chi_N^2}\right|_{\chi_N=\chi_S=0}.
\label{noise}
\end{eqnarray}
and similarly for higher order cumulants.

\section{Full counting statistics.}

In what follows we focus on the limit $kT,eV,E_{Th}\ll \Delta$. In
this limit, the physics is dominated by the induced proximity effect
in the diffusive region $N'$ between the two
barriers. \cite{proximity} The proximity effect affects the transport
properties on the energy scale $E_{Th}$. It can on a microscopic level
be described as resulting from coherence between the two
quasiparticles in the Andreev reflection, e.g. an incident electron
and a retroreflected hole. An induced proximity effect leads in
general to large modifications of the transport properties,
i.e. modifications scaling with the system size. We note that this is
in strong contrast to the situation in normal systems, where phase
coherence only gives rise to small quantum
corrections. \cite{Buttikerrew}

However, being a coherent phenomenon, the proximity effect is also
sensitive to modifications of the electron and hole phases. At finite
energies, the electron and hole pick up different phases and at $E \gg
E_{Th}$, the electron-hole coherence is destroyed and consequently,
the proximity effect is suppressed. Moreover, a magnetic flux in the
$N'$-region large enough to break time reversal symmetry, also
suppresses the proximity effect, since the electron and the hole pick
up different phases in a finite magnetic field. Here we however only
consider the case with negligibly small magnetic fields.

Formally, at energies well below the superconducting gap $\Delta$, the
matrix Greens function $\check G_S^0$ in Eq. (\ref{sres}) simplifies
considerably, since $\hat R_S=\hat A_S=\hat \sigma_x$ and consequently
$\hat K_S=0$. We then find $F(\chi_N,\chi_S)$ in Eq. (\ref{cgf}) by
diagonalizing $(\check G_x^2)^{1/2}$, giving
\begin{eqnarray}
F(\chi_N,\chi_S)=\frac{t_0}{e^2}\int \frac{dE}{2\sqrt{2}}\sqrt{\alpha+\sqrt{\alpha^2+\beta^2-4g_1^2g_2^2\Lambda}}, \nonumber \\ 
\label{cgflowenergy}
\end{eqnarray}
where we have introduced
\begin{eqnarray}
\alpha&=&g_1^2+g_2^2-\frac{E^2}{E_{Th}^2}(g_1+g_2)^2, \nonumber \\
\beta&=&2g_1(g_1+g_2)\frac{E}{E_{Th}}, \nonumber \\
\Lambda&=&\left(1-e^{2i(\chi_N-\chi_S)}\right)f_+(1-f_-)\nonumber \\
&+&\left(1-e^{-2i(\chi_N-\chi_S)}\right)f_-(1-f_+).
\label{countfield}
\end{eqnarray}
We note that at $E=0$, we recover the generating function of
Ref. [\onlinecite{Boerlin02}]. The terms $\mbox{exp}(\pm
2i[\chi_N-\chi_S])$ show that electrons are transported in pairs out
of ($+$) and into ($-$) the superconductor and into ($+$) and out of
($-$) the normal reservoir. This pair transport follows from the fact
that the charge transfer mechanism across the $NS$-interface is
Andreev reflection. The form of the cumulant generating function, with
the double square roots, tells that the pair transfer is highly
correlated. There are however several limits where the generating
function is simply proportional to $\Lambda$, describing an
uncorrelated, (generalized) Poissonian, transfer of pairs of electrons
(as follows from Eq. \ref{qcount}).

In the limit of high energy, $E \gg E_{Th}$, for arbitrary $g_1$ and
$g_2$, the integrand of the generating function in
Eq. (\ref{cgflowenergy}) takes on the form
\begin{eqnarray}
\frac{\partial F(\chi_N,\chi_S)}{\partial E}=\frac{t_0}{4e^2}\frac{g_1g_2^2}{(g_1+g_2)^2}\left(\frac{E_{Th}}{E}\right)^2\Lambda.
\label{cgfE}
\end{eqnarray}
For these energies, the probability of pair tunneling is strongly
suppressed and the pairs are thus emitted in a Poisson process. This
is a consequence of the different phases picked up by the electrons
and holes, $\sim \pm E/E_{Th}$, at finite energies. At $E \gg E_{Th}$
the electrons and holes loose their coherence and the pair tunneling
probability is suppressed.

In the limit of a dominating coupling of the diffusive region $N'$ to
the normal reservoir, $g_1\gg g_2$, the integrand of the generating
function becomes
\begin{eqnarray}
\frac{\partial F(\chi_N,\chi_S)}{\partial
E}=\frac{t_0}{4e^2}\frac{g_2^2}{g_1}\frac{1}{1+(E/E_{Th})^2}\Lambda.
\label{cgfg1g2}
\end{eqnarray}
Thus, in this limit, the pair transfer is also uncorrelated. In this
case, the barrier $I_1$ has a negligible resistance compared to $I_2$
and the junction is effectively a $NI_2S$-junction. As discussed
above, in this case the pair transfer is uncorrelated. We note that
for high energies $E \gg E_{Th}$, it coincides with Eq. (\ref{cgfE}),
(for $g_1\gg g_2$).
 
We also note\cite{Boerlin02} that for the limit $g_1 \ll g_2$, strong
coupling of $N'$ to the superconductor and for the additional limit of
low energies, $E \ll E_{Th}$, the pair transfer is uncorrelated. In
this case, the integrand of the generating function is simply
\begin{eqnarray}
\frac{\partial F(\chi_N,\chi_S)}{\partial
E}=\frac{t_0}{4e^2}\frac{g_1^2}{g_2}\Lambda
\label{cgfg2g1}
\end{eqnarray}
The reason for the uncorrelated pair transfer in this case is the
following: In the limit $g_1/g_2 \rightarrow 0$, a proximity gap opens
up in $N'$ and the junction becomes an effective $NI_1S'$ junction
($S'$ denoting the gapped normal region between the barriers). For
energies below the induced gap, the pair transfer is thus Poissonian,
as in a standard $NIS$-junctions. \cite{Khlus87,deJong94,Muz94}

Making a connection to the proximity effect, it can be noted that the
in two cases, Eq. (\ref{cgfE}) and Eq. (\ref{cgfg1g2}), the strongly
suppressed proximity effect in $N'$ is responsible for the
uncorrelated pair transport. However, in the last case,
Eq. (\ref{cgfg2g1}), the Poissonian pair transfer is due to the
induced proximity gap in $N'$.

It is instructive to also study qualitatively the full counting
statistics, starting from the scattering point of view in
Ref. [\onlinecite{Muz94}]. The cumulant generating function,
generalized to many modes, is
\begin{eqnarray}
F(\chi_N,\chi_S)=\frac{t_0}{h}\int~dE~\langle
\mbox{ln}(1+R_{eh}(E)\Lambda) \rangle
\label{cgfmuz}
\end{eqnarray}
where the $R_{eh}(E)$'s are the Andreev reflection eigenvalues,
i.e. the real and energy dependent eigenvalues of the hermitian
electron-hole scattering matrix product $S_{eh}^{\dagger}S_{eh}$ (see
e.g. Ref. [\onlinecite{deJong97}]). Here $\langle .. \rangle$ denotes
ensemble average over impurity configurations and $\Lambda$ is the
same as in Eq. (\ref{countfield}). For a general function $h(R_{eh})$,
the ensemble average can be written as an integration $\int dR_{eh}
\rho(R_{eh}) h(R_{eh})$, where $\rho(R_{eh})$ is the distribution of
Andreev reflection eigenvalues (the corresponding quantity in
NS-systems to the transmission eigenvalue distribution in normal
systems).

Expanding the logarithm in Eq. (\ref{cgfmuz}), makes it is clear that
the limits where the charge transfer is uncorrelated,
Eqs. (\ref{cgfE}) to (\ref{cgfg2g1}), corresponds to the limit where
$\langle R_{eh} \rangle \gg \langle R_{eh}^2 \rangle$ (and
consequently $\langle R_{eh} \rangle \gg \langle R_{eh}^n \rangle$ for
all higher moments $n>2$ since $0\leq R_{eh}\leq 1$).

From this we can draw the following conclusions. The limits of
uncorrelated pair transfer corresponds to the cases where there are no
or very few ``open'' Andreev channels, i.e with Andreev reflection
probability close to unity (leading to $\langle R_{eh} \rangle \gg
\langle R_{eh}^2 \rangle$). In the case where there are more open
channels, the various moments $\langle R_{eh}^n\rangle$ are of the
same order of magnitude, resulting e.g. in a noise below the
Poissonian noise of uncorrelated pair transfer across the
NS-interface, further discussed below. As is also further discussed
below, this Andreev channel picture provides a simple explanation for
the voltage and temperature dependence of the noise as well as the
current.

\section{Noise.}

The rest of the paper is devoted to a detailed analysis of the
properties of the the second cumulant, the noise, which is within
reach of existing experimental techniques. We note that the cumulant
generating function $F(\chi_N,\chi_S)$ in Eq. (\ref{cgflowenergy}) can
be written as $g_1$ (or $g_2$) times a function which only contains
the ratio $g_1/g_2$, i.e. all transport properties studied below,
(normalized) depend only on the ratio of the conductances.

\begin{figure}[h]        
\centerline{\psfig{figure=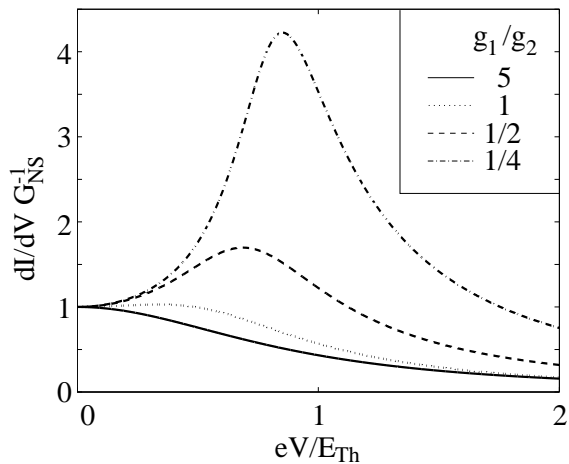,width=7.5cm}}
\caption{The differential conductance $dI/dV$, normalized with the
zero voltage conductance $G_{NS}$, as a function of $eV/E_{Th}$. The
values for the ratios $g_1/g_2=5$ and the corresponding line types are
shown in the legend. The temperature $kT \ll eV$. For $g_2=4g_1$, the
conductance shows a strong peak at $eV\sim E_{Th}$, a signature of an
ensemble averaged electron-hole resonance.}
\label{cond}
\end{figure}

For the benefit of the following discussion, we first study the
current, given from Eq. (\ref{curr}) and (\ref{cgflowenergy}),
\begin{eqnarray}
I=\frac{1}{e}\int
dE~\frac{f_--f_+}{\sqrt{2}}\frac{g_1^2g_2^2}{\sqrt{\alpha^2+\beta^2}\sqrt{\alpha+\sqrt{\alpha^2+\beta^2}}}
\label{currexp}
\end{eqnarray} 
The current has been studied both theoretically \cite{Volkov,Nazcirc}
and experimentally, \cite{Lefloch} here we briefly summarize the
findings. The differential conductance $G=dI/dV$ at zero temperature
is plotted as a function of $eV/E_{Th}$ in Fig. \ref{cond}, for
different ratios $g_1/g_2$.
The conductance for $eV\ll E_{Th}$ is given by
\begin{eqnarray}
G(eV\ll E_{Th})\equiv G_{NS}=\frac{g_1^2g_2^2}{(g_1^2+g_2^2)^{3/2}}.
\label{gnseq}
\end{eqnarray}
In the opposite limit, $eV\gg E_{Th}$, the conductance decays as a
power law with applied voltage, as
$(E_{th}/eV)^2g_2^2g_1/(g_1+g_2)^2$.

The behavior of the conductance for voltages of the order of $E_{Th}$
depends strongly on the ratio $g_1/g_2$. For $g_1\gg g_2$, i.e. for a
strong coupling of the normal region $N'$ to the normal reservoir, the
conductance decreases monotonically with voltage as
$(g_2^2/g_1)/[1+(eV/E_{Th})^2]$. In the opposite regime, $g_2\gg g_1$,
the normal region is strongly coupled to the superconducting
reservoir. The conductance has a strong peak at voltages $eV\sim
E_{Th}$, which can be seen as an ensemble averaged electron-hole
resonance. In other words, for energies $E\sim E_{Th}$ the
distribution of Andreev reflection eigenvalues, $\rho(R_{eh})$, is
shifted towards more ``open'' channels.

As pointed out above, in the limit of $g_1/g_2 \rightarrow 0$,
i.e. when decoupling the normal reservoir, a proximity induced gap
$E_{Th}$ opens up in the spectrum in the normal region $N'$. In this
limit, the conductance shows a singularity at $eV=E_{Th}$, similar to
the standard NIS-tunneling conductance, but with the induced gap
$E_{Th}$ instead of the superconducting gap $\Delta$.
\begin{figure}[h]        
\centerline{\psfig{figure=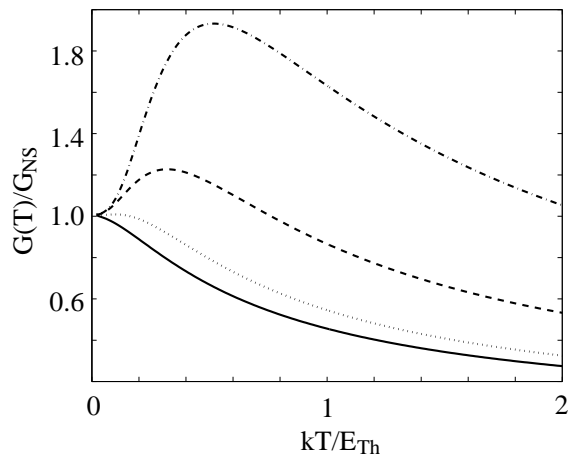,width=7.5cm}}
\caption{The zero voltage conductance normalized with $G_{NS}$, as a
function of temperature. The values for $g_1/g_2$ and the
corresponding line-types are the same as in Fig. \ref{cond}. The
conductance shows a temperature dependence which is qualitatively
similar to the voltage dependence of the conductance in
Fig. \ref{cond}. The zero voltage conductance is related to the
equilibrium noise as $S_{eq}(T)=4ekT G(eV=0,T)$.}
\label{tdepnoise}
\end{figure}

The temperature dependence of the zero voltage conductance, $G(T)$, is
shown in Fig. \ref{tdepnoise}. The temperature dependence is
qualitatively similar to the voltage dependence, shown in
Fig. \ref{cond}, due to the form of the spectral current density in
Eq. (\ref{currexp}). We note that for $kT \ll E_{Th}$ we have
$G=G_{NS}$ from Eq. (\ref{gnseq}). In the opposite limit, $kT \gg
E_{Th}$ the conductance decays as $1/kT$. In the intermediate regime,
the behavior of the conductance depends strongly on the ratio
$g_1/g_2$. For $g_1 \gg g_2$, it decays monotonically with
temperature, while in the opposite limit, $g_2 \gg g_1$, it has a peak
at temperatures $kT$ of order $E_{Th}$.

We then turn to the noise, found from Eq. (\ref{noise}) and
(\ref{cgflowenergy}) to be
\begin{eqnarray}
S&=&\int dE
\frac{2\sqrt{2}g_1^2g_2^2}{\sqrt{\alpha^2+\beta^2}\left(\alpha+\sqrt{\alpha^2+\beta^2}\right)^{3/2}}
\nonumber \\
&\times&\left[[f_+(1-f_-)+f_-(1-f_+)]\left(\alpha+\sqrt{\alpha^2+\beta^2}\right)\right. \nonumber \\
&&\left.-(f_--f_+)^2\frac{g_1^2g_2^2}{\alpha^2+\beta^2}
\left(2\alpha+3\sqrt{\alpha^2+\beta^2}\right)\right].
\label{noiseresult}
\end{eqnarray}
First we consider the equilibrium noise, for $eV\ll E_{Th},kT$. In
this limit we have $f_--f_+=0$ and the noise is
\begin{eqnarray}
S_{eq}&=&\int~dE~\frac{4\sqrt{2}f(1-f)g_1^2g_2^2}{\sqrt{\alpha^2+\beta^2}\sqrt{\alpha+\sqrt{\alpha^2+\beta^2}}}.
\label{eqnoise}
\end{eqnarray}
Noting that we have $2f(1-f)=kT[d(f_--f_+)/dV]_{eV=0}=-2kT(df/dE)$ we see from Eq. (\ref{currexp}) and (\ref{eqnoise}) that
\begin{eqnarray}
S_{eq}(T)&=&4kTG(eV=0,T),
\end{eqnarray}
for arbitrary temperature $T$. This is just what we expect from the
fluctuation-dissipation theorem. The temperature dependence of the
noise, normalized with $4kT$, is thus just the temperature dependence
of the zero voltage conductance, plotted in Fig. (\ref{tdepnoise}).

At $kT \ll E_{Th}$ we find just $S_{eq}(kT \ll E_{Th})=4kT G_{NS}$. In
the opposite limit, $kT \gg E_{Th}$ the equilibrium noise tends
towards a constant, temperature independent value
\begin{eqnarray}
S_{eq}(kT \gg E_{Th})&=&E_{Th}\frac{\pi g_1g_2^2}{(g_1+g_2)\sqrt{g_1^2+g_2^2}}.
\end{eqnarray}
In the intermediate temperature regime, the noise, as the conductance
in Fig. \ref{tdepnoise}, depends strongly on the ratio $g_1/g_2$.
 
\begin{figure}[h]        
\centerline{\psfig{figure=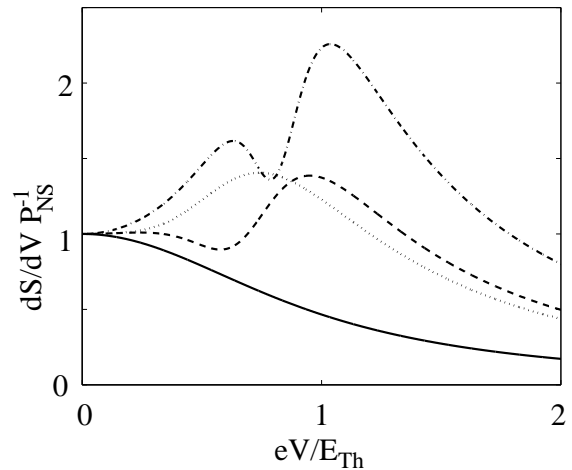,width=7.5cm}}
\caption{The differential shot noise $dS/dV$ as a function of applied
voltage, normalized with the zero voltage value. The values for
$g_1/g_2$ and the corresponding line-types are the same as in
Fig. \ref{cond}. The temperature $kT \ll E_{Th}$. The double-peak
behavior at $eV\sim E_{Th}$, for $g_2=4g_1$, is a signature of the
electron-hole resonance, responsible for the peak in the conductance
at $eV \sim E_{Th}$ in Fig. \ref{cond}.}
\label{vdepnoise}
\end{figure}

The voltage dependence of the noise is qualitatively different from
the temperature dependence. In the limit $kT \ll eV,E_{Th}$, the shot
noise regime, we have
\begin{eqnarray}
S&=&\int_{-eV}^{eV}~dE~\frac{\sqrt{2}g_1^2g_2^2}{\left(\alpha+\sqrt{\alpha^2+\beta^2}\right)^{3/2}}
\nonumber \\
&\times&\left(-1+\frac{3g_1^2g_2^2}{\alpha^2+\beta^2}-\frac{\alpha(\alpha^2+\beta^2-2g_1^2g_2^2)}{(\alpha^2+\beta^2)^{3/2}}\right).
\end{eqnarray}
Here we focus on the differential noise, $dS/dV$, which is plotted as
a function of voltage for different ratios $g_1/g_2$ in
Fig. \ref{vdepnoise}. 

In the low voltage limit, $eV \ll E_{Th}$, the differential noise is
given by \cite{deJong97,Boerlin02}
\begin{eqnarray}
\left.\frac{dS}{dV}\right|_{eV \ll E_{Th}}\equiv P_{NS}=2eG_{NS}\left(2-\frac{5g_1^2g_2^2}{(g_1^2+g_2^2)^2}\right).
\end{eqnarray}
The Fano factor, $P_{NS}/2eG_{NS}$, in this limit, is thus
$(2-5g_1^2g_2^2/[g_1^2+g_2^2]^2)$. It varies from $2$ in the limits
$g_1 \gg g_2$ and $g_2 \gg g_1$, to $3/4$ for\cite{Fauchere98}
$g_1=g_2$. A Fano factor $2$ is an indication of uncorrelated emission
of pairs of electrons into or out of the superconductor. As pointed
out in connection to Eq. (\ref{cgfg1g2}) and (\ref{cgfg2g1}), in these
limits, for $eV \ll E_{Th}$ and $g_1 \gg g_2$ or $g_2 \gg g_1$, the
generating function $F(\chi_N,\chi_S)$ in Eq. (\ref{cgflowenergy}) is
proportional to \cite{Boerlin02} $\mbox{exp}(2i[\chi_N-\chi_S])$,
showing that the charge transfer is indeed an uncorrelated emission
(for negative voltages) of pairs of electrons from the superconductor.

In the high voltage limit, $eV \gg E_{Th}$, the differential noise
decreases with voltage as
\begin{eqnarray}
\frac{1}{e}\frac{dS}{dV}=2\left(\frac{E_{th}}{eV}\right)^2
\frac{g_2^2g_1}{(g_1+g_2)^2}
\end{eqnarray}
i.e. the same power law behavior as the conductance, which is clear
directly from the generating function in Eq. (\ref{cgfE}).

At intermediate voltages, the noise depends strongly on the ratio
$g_1/g_2$. For a dominating coupling of the normal region to the
normal reservoir, $g_1\gg g_2$, the differential noise decreases
monotonically with applied voltage, as
\begin{eqnarray}
\frac{1}{e}\frac{dS}{dV}=2\frac{g_2^2}{g_1}\frac{1}{1+(eV/E_{Th})^2}. 
\end{eqnarray}
This can also be obtained directly from the generating function in
Eq. (\ref{cgfg1g2}). We note that this behavior is very similar to
what was found for a diffusive $NIS$-junction,\cite{Hekkila02} a
diffusive normal region of non-negligible resistance, connected to a
superconducting reservoir via a tunnel barrier.

In the opposite limit, $g_1\gg g_2$, the noise shows a double peak
structure around $eV\sim E_{Th}$, the voltage where the conductance
shows a single peak (see Fig. \ref{cond}). Such a double peak behavior
has been discussed before in single mode junctions with sharp Andreev
resonances. \cite{Fauchere98} In single mode junctions, the double
peak behavior follows directly from the fact that if the Andreev
reflection probability $R_{eh}$ has a resonance at some energy $E_0$,
the noise, $\sim R_{eh}(1-R_{eh})$, shows a double peak around
$E_0$. Interestingly, as is clear from Fig. \ref{vdepnoise}, in the
$NI_1N'I_2S$, this double-peak behavior survives the ensemble
average. This indicates that the whole distribution $\rho(R_{eh})$ of
Andreev reflection eigenvalues is shifted from being dominated by
``closed'' towards being dominated by ``open'' channels at resonance
$E \sim E_{Th}$.

We note that such a double peak behavior can also be found in normal
single-mode double-barrier junctions, it is not an effect of the
superconductivity (see Ref. [\onlinecite{Buttikerrew}]). However, in
normal, many-mode systems, such a behaviour is, to leading order in
number of modes, washed out when performing an ensemble average.
\begin{figure}[h]        
\centerline{\psfig{figure=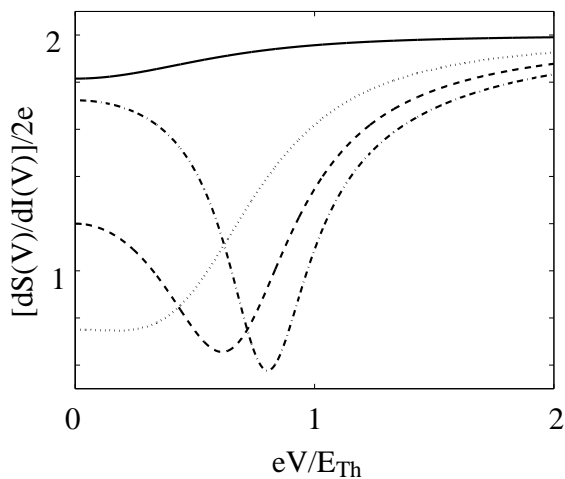,width=7.5cm}}
\caption{The differential Fano factor $[dS(V)/dI(V)]/2e$ as a function
of applied voltage. The values for $g_1/g_2$ and the corresponding
line-types are the same as in Fig. \ref{cond}. The temperature $kT \ll
E_{Th}$. The dip at $eV\sim E_{Th}$, for $g_2=4g_1$, is a signature of
the electron-hole resonance, responsible for the peak in the
conductance at $eV \sim E_{Th}$ in Fig. \ref{cond}.}
\label{vdepfano}
\end{figure}

Additional insight can be obtained by studying the differential Fano
factor, $[dS(V)/dI(V)]/2e$. In the differential Fano factor, the
voltage dependence of the conductance is canceled out from the noise,
and we obtain the ``bare'' voltage-dependence of the noise. We have
\begin{eqnarray}
\frac{1}{2e}\frac{dS(V)}{dI(V)}&=&2-\frac{2g_1^2g_2^2\left(2\alpha+3\sqrt{\alpha^2+\beta^2}\right)}{(\alpha^2+\beta^2)\left(\alpha+\sqrt{\alpha^2+\beta^2}\right)}.
\end{eqnarray}
This quantity is plotted in Fig. \ref{vdepfano}. We note that the
differential Fano factor shows a much stronger dependence of the
applied voltage, compared to earlier studied
\cite{Belzig01a,Reulet02,Hekkila02,Belzig02b,Reulet02b} diffusive $NS$
and $NIS$ junctions. The resonant double-peak in the noise, for $g_2
\gg g_1$ is not visible in the differential Fano factor, instead it
shows a dip at $eV\sim E_{Th}$. This is in accordance with our
explanation of the double peak-behavior of the noise above, since,
again making the comparison to a single mode junction, the
differential Fano factor is the differential shot noise divided by the
differential conductance, $\sim [R_{eh}(1-R_{eh})]/R_{eh}=1-R_{eh}$,
which thus has a dip at the resonant energy.

Moreover, in the high voltage limit, $eV \gg E_{Th}$, the differential
Fano factor saturates at a constant value $2$, i.e. independent on the
relation between the conductances $g_1$ and $g_2$. This is a
manifestation of the uncorrelated charge transfer at these energies,
as discussed in connection to Eq. (\ref{cgfE}). Also in the limit $g_2
\gg g_1$, the Fano factor approaches $2$, independent on voltage. This
is again a manifestation of the uncorrelated pair transfer, and can be
seen directly from the generating functions in
Eq. (\ref{cgfg1g2}). From this we conclude that the differential Fano
factor is thus the relevant quantity for studying the energy
dependence of the charge transfer mechanism.

\section{Conclusions}

In conclusion, we have investigated the energy dependent full counting
statistics of a diffusive
normal-insulating-normal-insulating-superconducting ($NI_1N'I_2S$)
junction. We have used the recently developed circuit theory of full
counting statistics, allowing us to access the full temperature and
voltage dependence of the statistics. In general, the charge is
transported via correlated transfer of pairs of electrons. However, in
the case of strongly asymmetric contacts or energies much larger than
the Thouless energy, the pair transfer is uncorrelated. The second
cumulant, the noise, was studied in detail. It was found to depend
strongly on voltage and temperature. For low temperatures and a
junction resistance dominated by the tunnel barrier to the normal
reservoir ($I_1$), the noise shows a double-peak behavior at voltages
$eV\sim E_{Th}$, a signature of an ensemble averaged electron-hole
resonance.

We acknowledge discussions with F. Lefloch, W. Belzig and
M. B\"uttiker. This work was supported by the Swiss National Science
Foundation and the project for Materials with Novel Electronic
Properties.


\begin{thebibliography}{999}
\bibitem{relevant} 
We always have e.g. the solution $\check G=1$,
which has no physical meaning.
\end{thebibliography}
\end{document}